# Variable Selection in Regression Model with AR(p) Error Terms Based on Heavy Tailed Distributions

Y. Tuaç[1] and O. Arslan[1]

[1]Ankara University, Faculty of Science, Department of Statistics, 06100 Ankara/Turkey
ytuac@ankara.edu.tr  oarslan@ankara.edu.tr

**Abstract**

Parameter estimation and the variable selection are two pioneer issues in regression analysis. While traditional variable selection methods require prior estimation of the model parameters, the penalized methods simultaneously carry on parameter estimation and variable select. Therefore, penalized variable selection methods are of great interest and have been extensively studied in literature. However, most of the papers in literature are only limited to the regression models with uncorrelated error terms and normality assumption. In this study, we combine the parameter estimation and the variable selection in regression models with autoregressive error term by using different penalty functions under heavy tailed error distribution assumption. We conduct a simulation study and a real data example to show the performance of the estimators.

**JEL Codes:** C150, C510, C520.

**Keywords:** *Dependent errors, EM algorithm*, LASSO, *heavy tailed distributions, variable selection.*

## 1. Introduction

The high dimensional nature of the existing data sets is required new variable section and parameter estimation methods which has brought great attention to the statisticians. Variable selection focuses on searching for the best set of relevant variables to include in a model. In linear regression models there are two popular variable selection procedures: subset selection and penalized methods. Although, subset selection methods are quite practically useful, these methods need parameter estimation before performing the variable selection and they often show high variability which cannot reduce the prediction error of the full model. On the other hand, penalized regression methods which we consider in this study such as LASSO, bridge, ridge, SCAD and elastic-net are select variables and estimate model parameters simultaneously. Hence, these methods provide less prediction error and more stability than the subset selection methods. We can briefly summarize the basic properties of these methods as follows: Least Absolute Shrinkage Selection Operator (LASSO) estimation method (Tibshirani, 1996) the OLS loss function $(y - X\beta)^T(y - X\beta)$ is minimized under the restriction $\sum_{j=1}^{p} |\beta_j| \leq t$. LASSO use $L_1$ type penalized function. Unlike ridge regression, there is no analytic solution for the LASSO therefore numeric methods should be used. LASSO estimation method sets the unimportant coefficients to zero to achieve the model selection while doing the parameter estimation simultaneously. Smoothly Clipped Absolute Deviation (SCAD) introduced by Fan and Li (2001) possesses the oracle properties unlike the LASSO such as: unbiasedness for large true coefficient to avoid excessive estimation bias, sparsity to reduce model complexity, continuity to avoid unnecessary variation in model prediction. According to Fan and Li (2001) optimal $\lambda$ can be choose by using Cross Validation (CV) method, it does not bring any difference to select $\alpha$ parameter. Fan and Li (2001) suggested taking $\alpha = 3.7$ is a good choice for various problem, we also use the same value in simulation study. Bridge, introduced by Frank and Friedman (1993) utilizes the $L_\gamma$ ($\gamma > 0$) penalty and thus, it includes the ridge (Hoerl and Kennard, 1970) ($\gamma = 2$) and the LASSO ($\gamma = 1$) as special cases. Bridge estimators produce sparse models when $0 < \gamma \leq 1$. In the literature there are some algorithms which proposed to choose optimal value of $\gamma$ (Park and Yoon, 2011, Liu et al., 2007). In this study we choose $\gamma$ and $\lambda$ parameters by using a two-dimensional grid search. If some of the covariates are highly correlated, then the elastic-net proposed by Zou and Hastie (2005) which combines $L_1$ and $L_2$ norms can be used.

These methods are widely used in literature and most studies have been done under assumption that observations are independent. However, if the error terms are correlated in time and ignoring this correlation effect while estimating the model parameters may result in incorrectly estimate of the variance of the model parameters. This causes the invalid resulting of confidence intervals and hypothesis tests. To overcome this problem, we assumed p order autoregressive (AR(p)) (Box and Jenkins, 1976) structure to model the error terms. This type of regression models has been considered before both parameter estimation without variable selection and penalized based variable selection cases.

Concerning the parameter estimation, some papers in literature which use normal distribution as error distribution assumption might be listed as follows. Alpuim and El-Shaarawi (2008) estimate parameters of the regression model with AR(p) error term using the ordinary least square (OLS) estimation method. They also use the maximum likelihood (ML) and conditional maximum likelihood (CML) estimation methods under the assumption of normality and study the asymptotic properties of the resulting estimators. Beach and Mackinnon (1978) used ML method to estimate the parameters of AR(1) error term regression models. However, normality assumption is very restrictive and sensitive to the existence of outliers in the data. The *t* distribution provides a useful alternative to the normal distribution for statistical modelling of data sets that have heavier tailed empirical distribution. Tiku (2007) estimates the parameters by using the modified maximum likelihood (MML) method for the regression model with AR(1) error terms under the assumption that the error term has heavy tailed distribution. Recently, Tuaç et al. (2018) proposed to use symmetric *t* distribution as the distribution for error terms and used CML method to carry on the estimation of the model parameters in autoregressive error term regression model as a robust alternative to the normal distribution. Further, Tuaç et al. (2020) used skew alternatives to the normal and the *t* distributions as error distributions in AR(p) error term regression model.

Regarding the penalized variable selection case there are also some studies are exist based on AR(p) error term regression model. Wang et al. (2007) consider linear regression with autoregressive errors. They employ the modified LASSO-type penalty both on regression and autoregressive coefficients. Yoon et al. (2012) consider adaptive LASSO, SCAD and bridge method in AR(p) error term regression models under normality assumption. However, these papers are not considered any other distribution apart from the normal. In the regression model, the error terms may not always conform to the normal distribution in real life. The distribution of the data may be thicker-tailed than the normal distribution. Using normal distribution for the error terms might result lack of the performance of the selection of the important variables in the model besides the higher MSE (mean squared error) of the parameter estimations. To solve this problem, as an alternative to the normal distribution, the symmetric *t* distribution, which exhibits a thicker-tailed behavior than the normal distribution, is considered (Tuaç, 2020). Using *t* distribution as error distribution improve the ability of the selection of the important variables and the resulting smaller MSE of the parameter estimations when there are outliers exist. In this paper, we construct five different penalized regression methods; LASSO, bridge, ridge, SCAD and elastic-net in the autoregressive error term regression model with the distribution assumption of *t* distribution as a heavy tailed alternative to the normal distribution. To handle the complex maximization problem, we combine the Expectation Maximization (EM) (Dempster et.al, 1977, McLachilan and Krishnan, 1997) algorithm with aforementioned variable selection methods.

The remainder of the paper is organized as follows, Section 2 gives concise information about LASSO, bridge, SCAD and elastic-net estimators. In Section 3, we discuss penalizing the autoregressive error term regression model with mentioned methods. In Section 4, we give some brief description of EM algorithm and we propose an Expectation Conditional Maximization (ECM) (Meng, X. L. and Rubin, D. B., 1993) type of algorithm to calculate the ML estimate of the parameters and variable selection of penalized regression model with AR(p) error terms for each five methods under *t* distribution assumption. Section 5 presents a simulation study and a real data example for comparison of the five methods based on *t* distribution. Finally, Section 6 summarize the paper with a conclusion.

## 2. Penalized variable selection methods

Consider the following linear regression model

$$y_t = \sum_{i=1}^{M} x_{t,i} \beta_i + e_t, \qquad t = 1, 2, \ldots, N \tag{1}$$

where, $y_t$ is the response variable, $x_{t,i}$ explanatory variables, $\beta_i$ regression parameters and $e_t$ are independent and identically distributed random variables with mean 0 and variance $\sigma^2$. Penalized regression methods can be done by maximizing the following objective function

$$L_N = -\ln L(\boldsymbol{\beta}, \sigma^2 | e_t) + N \sum_{i=1}^{p} p_{\lambda_i}(\beta_i), \tag{2}$$

where $lnL(.)$ is the log-likelihood function and $p_{\lambda_i}(.)$ is a penalty function and $\lambda_i$'s are penalty parameters.

In this paper, we consider five different forms of $p_{\lambda_i}(.)$:

(i) $p_\lambda^{LASSO}(\beta_j) = \lambda \sum_{j=1}^{p} |\beta_j|, \quad \lambda > 0$

(ii) $p_\lambda^{SCAD}(\beta_j) = \begin{cases} \lambda|\beta_j| & , \quad |\beta_j| \leq \lambda \\ -\left(\dfrac{|\beta_j|^2 - 2\alpha\lambda|\beta_j| + \lambda^2}{2(\alpha-1)}\right) & , \quad \lambda < |\beta_j| \leq \alpha\lambda \\ \dfrac{(\alpha+1)\lambda^2}{2} & , \quad |\beta_j| > \alpha\lambda \end{cases}$

(iii) $p_\lambda^{Bridge}(\beta_j) = \lambda \sum_{j=1}^{p} |\beta_j|^\gamma , \quad 0 < \gamma \leq 1$

(iv) $p_\lambda^{Ridge}(\beta_j) = \lambda \sum_{j=1}^{p} \beta_j^2 , \quad \lambda > 0$

(v) $p_\lambda^{Elastic-Net}(\beta_j) = \lambda_1 \sum_{j=1}^{p} |\beta_j| + \lambda_2 \sum_{j=1}^{p} \beta_j^2$

## 3. Estimation in penalized autoregressive regression model

We consider model (1) with error terms have dependent structure as AR(p) model as follows

$$a_t = e_t - \phi_1 e_{t-1} - \cdots - \phi_p e_{t-p}, \quad |\phi_j| < 1 , \quad j = 1,2,\ldots,p . \tag{3}$$

For simplicity, following backshift operator $(B)$ can be used

$$a_t = \Phi(B)e_t \tag{4}$$

where $a_t$ will be the new error term for regression model. $E(a_t) = 0, Var(a_t) = \sigma^2$ and they are uncorrelated random variables with constant variance. Then, with the help of backshift operator, the regression model given in (1) with autoregressive error terms can be rewritten as

$$\Phi(B)y_t = \sum_{i=1}^{M} \beta_i \Phi(B)x_{t,i} + a_t, \quad t = 1,2,\ldots,T \tag{5}$$

where,
$\Phi(B)y_t = y_t - \phi_1 y_{t-1} - \cdots - \phi_p y_{t-p}$,
$\Phi(B)x_{t,i} = x_{t,i} - \phi_1 x_{t-1,i} - \cdots - \phi_p x_{t-p,i}$,
$M$: number of regression coefficients,
$T = N - p$,
$p$: autoregressive degree,
$N$: sample size,
$\phi_j$ are the AR(p) model parameters for $j = 1,2,\ldots,p$ and $a_t$ iid error terms with white noise process.

## 3.1 Normal distribution assumption

The parameter estimations and variable selection of regression model given in (4) can be done with penalized objective function as below. (Yoon et al., 2012)

$$Q_N(\boldsymbol{\beta}, \boldsymbol{\phi}) = \sum_{t=p+1}^{N}\left(y_t - \boldsymbol{x}_t^T\boldsymbol{\beta} - \sum_{j=1}^{p}\phi_j(y_{t-j} - \boldsymbol{x}_{t-j}^T\boldsymbol{\beta})\right)^2 + T\sum_{i=1}^{M} p_{\lambda_i}|\beta_i| + T\sum_{j=1}^{p} p_{\delta_j}|\phi_j| \quad (6)$$

Under the assumptions that $a_t \sim N(0, \sigma^2)$, this function also equivalently can be used as negative log likelihood function with penalized term as follows.

$$Q(\boldsymbol{\beta}, \boldsymbol{\phi}, \sigma^2) = -\ln L_N(\boldsymbol{\beta}, \boldsymbol{\phi}, \sigma^2|a_t) + T\sum_{i=1}^{M} p_{\lambda_i}|\beta_i| + T\sum_{j=1}^{p} p_{\delta_j}|\phi_j| \quad (7)$$

where

$p_{\lambda_i}|\beta_i|$: penalize function for $\beta$,
$p_{\delta_j}|\phi_j|$: penalize function for $\phi$,

and

$$\ln L(\boldsymbol{\beta}, \boldsymbol{\phi}, \sigma^2|a_t) = c - \frac{T}{2}\ln(\sigma^2) - \frac{1}{2\sigma^2}\sum_{t=p+1}^{N}\left(\Phi(B)y_t - \sum_{i=1}^{M}\beta_i \Phi(B)x_{t,i}\right)^2 \quad (8)$$

is the log likelihood function of the normal distribution. Someone can replace $p_{\lambda_i}|\beta_i|$ function with interested penalized methods penalize function then the parameter estimation and the variable selection can be done simultaneously by minimizing equation (6) or maximizing equation (7). Note that according to Yoon et al. (2012), sake of the ease of the computation, penalizing the autoregressive parameters are ignored. In this study, we also calculate both regression and autoregression parameter estimations with only penalizing the regression model parameters.

The penalized functions belong to LASSO, SCAD, bridge and elastic-net methods are not differentiable at the origin for the minimization problem in equation (6) and because they are not concave, Local Quadratic Approximation (LQA) which is proposed by Fun and Li (2001) can be used to solve this problem. According to the LQA the penalized functions are locally approximated at $\boldsymbol{\beta}^{(0)} = (\beta_1^0, \dots, \beta_M^0)$ in the following quadratic function

$$p_{\lambda_i}(|\beta_i|) \approx p_{\lambda_i}\left(\left|\beta_i^{(0)}\right|\right) + \frac{1}{2}\frac{p'_{\lambda_i}\left(\left|\beta_i^{(0)}\right|\right)}{\left|\beta_i^{(0)}\right|}\left(\beta_i^2 - \beta_i^{(0)2}\right)$$

$$= \frac{1}{2}\frac{p'_\lambda\left(\left|\beta_i^{(0)}\right|\right)}{\left|\beta_i^{(0)}\right|}\beta_i^2 + p_\lambda\left(\left|\beta_i^{(0)}\right|\right) - \frac{1}{2}\frac{p'_\lambda\left(\left|\beta_i^{(0)}\right|\right)}{\left|\beta_i^{(0)}\right|}\beta_i^{(0)2} \quad (9)$$

where,

$p'_{\lambda_i}$ : derivative of the interested penalized function
$\beta_i^{(0)}$: initial values of $\beta_i$

By using this LQA method the minimization problem in equation (9) will be reduced the following objective function around $\beta^{(0)}$.

$$Q(\boldsymbol{\beta}, \boldsymbol{\phi}, \sigma^2) = \sum_{t=p+1}^{N} \left( y_t - x_t^T \boldsymbol{\beta} - \sum_{j=1}^{p} \phi_j (y_{t-j} - x_{t-j}^T \boldsymbol{\beta}) \right)^2 + \frac{T}{2} \sum_{i=1}^{M} \frac{p'_{\lambda_i}\left(\left|\beta_i^{(0)}\right|\right)}{\left|\beta_i^{(0)}\right|} \beta_i^2 \qquad (10)$$

By minimizing the above objective function the parameter estimations and the variable selections for autoregressive error term regression model with *t* distribution assumption can be made simultaneously.

$$\widehat{\boldsymbol{\beta}} = \left[\widehat{\Phi}(B) X^T \widehat{\Phi}(B) X + T p_\lambda(\boldsymbol{\Omega})\right]^{-1} \left[\widehat{\Phi}(B) X^T \widehat{\Phi}(B) Y\right] \qquad (11)$$

where

$p_\lambda(\boldsymbol{\Omega})$: is the diagonal matrix derived from LQA method with penalty parameter of interested variable selection method.

$$\widehat{\boldsymbol{\phi}} = \mathbf{R}(\widehat{\boldsymbol{\beta}})^{-1} R_0(\widehat{\boldsymbol{\beta}}) \qquad (12)$$

where

$$R_0(\widehat{\boldsymbol{\beta}}) = \begin{bmatrix} \sum_{t=p+1}^{N} e_t e_{t-1} \\ \sum_{t=p+1}^{N} e_t e_{t-2} \\ \vdots \\ \sum_{t=p+1}^{N} e_t e_{t-p} \end{bmatrix}, \mathbf{R}(\widehat{\boldsymbol{\beta}}) = \begin{bmatrix} \sum_{t=p+1}^{N} e_{t-1}^2 & \sum_{t=p+1}^{N} e_{t-1} e_{t-2} & \cdots & \sum_{t=p+1}^{N} e_{t-1} e_{t-p} \\ \sum_{t=p+1}^{N} e_{t-2} e_{t-1} & \sum_{t=p+1}^{N} e_{t-2}^2 & \cdots & \sum_{t=p+1}^{N} e_{t-2} e_{t-p} \\ \vdots & \vdots & \ddots & \vdots \\ \sum_{t=p+1}^{N} e_{t-p} e_{t-1} & \sum_{t=p+1}^{N} e_{t-p} e_{t-2} & \cdots & \sum_{t=p+1}^{N} e_{t-p}^2 \end{bmatrix}.$$

$$\widehat{\sigma}^2 = \frac{1}{T} \sum_{t=p+1}^{N} \left(\widehat{\Phi}(B) y_t - \widehat{\Phi}(B) x_t \widehat{\boldsymbol{\beta}}\right)^2 \qquad (13)$$

### 3.2 *t* distribution assumption

In this paper, we use ECM algorithm to find the CML estimates. Note that when *t* distribution is considered in different models, the ECM algorithm is extensively used because of the normal mixture representation of the $t$ distribution. In the next section, we show how we employ the ECM algorithm to estimate the AR(p) regression model parameters under $t$ distribution assumption. The probability density function (pdf) of Student's $t$ distribution which we used in this article is as follow.

$$f(x) = \frac{c_v}{\sigma}\left(v + \frac{x^2}{\sigma^2}\right)^{-\frac{v+1}{2}}, \qquad -\infty < x < \infty \qquad (14)$$

where $c_v = \frac{\Gamma\left(\frac{v+1}{2}\right)v^{v/2}}{\sqrt{\pi}\Gamma\left(\frac{v}{2}\right)}$, $v > 0$ degrees of freedom and $\sigma^2 > 0$ scale parameter.

The $t$ distribution is a symmetric heavy tailed alternative to the normal distribution and it is more resistant than the normal distribution to the outliers. Consider the model in equation (5) and assume that error terms have the $t$ distribution $(a_t \sim t_v(0, \sigma^2, v))$. The CML estimators of the unknown parameters $\boldsymbol{\theta} = (\boldsymbol{\beta}, \boldsymbol{\phi}, \sigma^2, v)$ can be obtained by using the following conditional log likelihood function

$$\ln L(\boldsymbol{\theta}|a_t) = \ln c_v - T \ln \sigma - \frac{v+1}{2} \sum_{t=p+1}^{N} \ln\left[v + \frac{\left(\Phi(B)y_t - \sum_{i=1}^{M}\beta_i \Phi(B)x_{t,i}\right)^2}{\sigma^2}\right]. \qquad (15)$$

The estimators of the parameter vector $\boldsymbol{\theta}$ can be obtained by maximizing this conditional log likelihood function. However, it is not tractable to obtain the maximum of this function. Therefore, some numerical methods should be used to obtain the CML estimates of the parameters. Among these numerical methods, the ECM algorithm will be slightly easy to use because of the normal mixture representation of the $t$ distribution. First, hierarchical formulation in terms of the conditional distributions can be expressed as follows.

$$\Phi(B)y_t|u_t \sim N\left(\Phi(B)x_t^T\boldsymbol{\beta}, \frac{\sigma^2}{u_t}\right), \qquad (16)$$
$$u_t \sim Gamma\left(\frac{v}{2}, \frac{v}{2}\right),$$

where $\boldsymbol{u} = (u_1, \ldots, u_n)$ is missing and $y$ is the observed data, respectively. Then the conditional pdf is

$$u_t|\Phi(B)y_t \sim Gamma\left(\frac{v+1}{2}, \frac{1}{2}(v + \kappa_t^2)\right), \qquad (17)$$

where $\kappa_t^2 = \frac{\left(\Phi(B)y_t - \Phi(B)x_t^T\boldsymbol{\beta}\right)^2}{\sigma^2}$. Considering $u$ is missing data and the complete data log likelihood function for $(\boldsymbol{y}, \boldsymbol{u})$ can be written as

$$l_c(\boldsymbol{\theta}; \boldsymbol{y}, \boldsymbol{u}) = -\frac{T}{2}\ln(2\pi\sigma^2) + \frac{1}{2}\sum_{t=p+1}^{N} \ln u_t$$

$$-\sum_{t=p+1}^{N}\left(\frac{\kappa_t^2}{2}+\frac{v}{2}\right)u_t+\left(\frac{v}{2}-1\right)\sum_{t=p+1}^{N}\ln u_t+T\ln c_v \qquad (18)$$

According to the procedure of the ECM algorithm, the conditional expectation of the complete data log likelihood function, given the observed data and the current parameter estimate should be calculated. Therefore, the following conditional expectation should be calculated.

$$E(l_c(\boldsymbol{\theta};\boldsymbol{y},\boldsymbol{u})|\Phi(B)y_t) = -\frac{T}{2}\ln(2\pi\sigma^2) + T\ln c_v$$

$$+\frac{1}{2}\sum_{t=p+1}^{N} E(\ln u_t\,|\Phi(B)y_t)$$

$$-\sum_{t=p+1}^{N}\frac{\kappa_t^2}{2}E(u_t|\Phi(B)y_t)$$

$$-\frac{v}{2}\sum_{t=p+1}^{N} E(u_t|\Phi(B)y_t)$$

$$+\left(\frac{v}{2}-1\right)\sum_{t=p+1}^{N} E(\ln u_t\,|\Phi(B)y_t) \qquad (19)$$

The conditional expectations $E(u_t|\Phi(B)y_t)$ and $E(\ln u_t\,|\Phi(B)y_t)$ can be computed as follows.

$$E(u_t|\Phi(B)y_t) = \frac{v+1}{v+\kappa_t^2}, \qquad (20)$$

$$E(\ln u_t\,|\Phi(B)y_t) = DG\left(\frac{v+1}{2}\right) - \ln\left(\frac{1}{2}(v+\kappa_t^2)\right), \qquad (21)$$

where $DG$ is digamma function. Then the steps of the ECM algorithm can be given as follows.

*The steps of the ECM algorithm:*
The steps of the ECM algorithm for this study can be briefly summarized as follows.
***E step*** Find the conditional expectations of the complete data log likelihood function given $y_t$ and current parameter estimates $\boldsymbol{\theta}^{(k)}$. These conditional expectations result the objective function $Q(\boldsymbol{\theta};\boldsymbol{\theta}^{(k)})$ to be maximized.
***M step*** Maximize the objective function $Q(\boldsymbol{\theta};\boldsymbol{\theta}^{(k)})$ with respect to the parameters $\boldsymbol{\theta}$ to obtain the new estimates $\boldsymbol{\theta}^{(k+1)}$.
These steps are very general. For the details of the algorithm to complete the estimates to following IRA can be given for the *t* distribution.

1. Determine the initial value $\boldsymbol{\theta}^{(1)} = \left(\boldsymbol{\beta}^{(1)}, \boldsymbol{\phi}^{(1)}, \sigma^{2(1)}, \nu^{(1)}\right)$ of the parameter vector $\boldsymbol{\theta}$ and stopping criteria $\eta$.

2. *(E-step)* Compute the following conditional expectations using the current estimates $\boldsymbol{\theta}^{(k)} = \left(\boldsymbol{\beta}^{(k)}, \boldsymbol{\phi}^{(k)}, \sigma^{2(k)}, \nu^{(k)}\right)$ for $k = 1,2,3 \ldots$

$$\hat{S}_{1t}^{(k)} = E\left(u_t \middle| \Phi(B)y_t, \widehat{\boldsymbol{\theta}}^{(k)}\right) = \frac{\nu^{(k)} + 1}{\nu^{(k)} + \kappa_t^{(k)2}} \qquad (22)$$

$$\hat{S}_{2t}^{(k)} = E\left(\ln u_t \middle| \Phi(B)y_t, \widehat{\boldsymbol{\theta}}^{(k)}\right) = DG\left(\frac{\nu^{(k)} + 1}{2}\right) - \ln\left(\frac{1}{2}\left(\nu^{(k)} + \kappa_t^{(k)2}\right)\right) \qquad (23)$$

3. *(E-step)* Use these conditional expectations in the objective function $Q(\boldsymbol{\theta}; \widehat{\boldsymbol{\theta}})$ to form the following function to be maximized with respect to $\boldsymbol{\theta}$. By writing these conditional expectations in equation (21) the objective function to be maximized at M step can be obtained as follows:

$$Q(\boldsymbol{\theta}; \widehat{\boldsymbol{\theta}}^{(k)}) = -\frac{T}{2}\ln(2\pi\sigma^2) + T\ln(c_\nu) + \frac{1}{2}\sum_{t=p+1}^{N} \hat{S}_{2t}^{(k)} - \sum_{t=p+1}^{N} \frac{\kappa_t^2}{2}\hat{S}_{1t}^{(k)}$$

$$-\frac{\nu}{2}\sum_{t=p+1}^{N} \hat{S}_{1t}^{(k)} + \left(\frac{\nu}{2} - 1\right)\sum_{t=p+1}^{N} \hat{S}_{2t}^{(k)} - \frac{T\lambda}{2}\sum_{i=1}^{M} \left|\beta_i^{(0)}\right|^{-1}\beta_i^2 \qquad (24)$$

4. *(CM step)* Maximize $Q(\boldsymbol{\theta}; \widehat{\boldsymbol{\theta}}^{(k)})$ with respect to $\boldsymbol{\theta} = (\boldsymbol{\beta}, \boldsymbol{\phi}, \sigma^2, \nu)$ to obtain $\boldsymbol{\theta}^{(k+1)} = \left(\widehat{\boldsymbol{\beta}}^{(k+1)}, \widehat{\boldsymbol{\phi}}^{(k+1)}, \hat{\sigma}^{2(k+1)}, \nu^{(k+1)}\right)$. By doing this we obtain following estimates at *(k+1)* th iteration

$$\widehat{\boldsymbol{\beta}} = \left[\hat{S}_{1t}^{(k)}\widehat{\Phi}(B)X^T\widehat{\Phi}(B)X + Tp_\lambda(\boldsymbol{\Omega})\right]^{-1}\left[\hat{S}_{1t}^{(k)}\widehat{\Phi}(B)X^T\boldsymbol{\omega}\widehat{\Phi}(B)Y\right] \qquad (25)$$

$$\widehat{\boldsymbol{\phi}}^{(k+1)} = (R_s)^{-1}(R_{0s}) \qquad (26)$$

$$\hat{\sigma}^{2(k+1)} = \frac{1}{T}\left(\sum_{t=p+1}^{N} \hat{S}_{1t}^{(k)}\left(\widehat{\Phi}(B)y_t - \widehat{\Phi}(B)x_t^T\widehat{\boldsymbol{\beta}}^{(k)}\right)^2\right) \qquad (27)$$

$p_\lambda(\boldsymbol{\Omega})$: is the diagonal matrix derived from LQA method with penalty parameter of interested variable selection method.
Here

$$\mathbf{R}_{0s} = \begin{bmatrix} \sum_{t=p+1}^{N} \hat{S}_{1t}^{(k)} e_t e_{t-1} \\ \sum_{t=p+1}^{N} \hat{S}_{1t}^{(k)} e_t e_{t-2} \\ \vdots \\ \sum_{t=p+1}^{N} \hat{S}_{1t}^{(k)} e_t e_{t-p} \end{bmatrix}, \mathbf{R}_s = \begin{bmatrix} \sum_{t=p+1}^{N} \hat{S}_{1t}^{(k)} e_{t-1}^2 & \sum_{t=p+1}^{N} \hat{S}_{1t}^{(k)} e_{t-1} e_{t-2} & \cdots & \sum_{t=p+1}^{N} \hat{S}_{1t}^{(k)} e_{t-1} e_{t-p} \\ \sum_{t=p+1}^{N} \hat{S}_{1t}^{(k)} e_{t-2} e_{t-1} & \sum_{t=p+1}^{N} \hat{S}_{1t}^{(k)} e_{t-2}^2 & \cdots & \sum_{t=p+1}^{N} \hat{S}_{1t}^{(k)} e_{t-2} e_{t-p} \\ \vdots & \vdots & \ddots & \vdots \\ \sum_{t=p+1}^{N} \hat{S}_{1t}^{(k)} e_{t-p} e_{t-1} & \sum_{t=p+1}^{N} \hat{S}_{1t}^{(k)} e_{t-p} e_{t-2} & \cdots & \sum_{t=p+1}^{N} \hat{S}_{1t}^{(k)} e_{t-p}^2 \end{bmatrix}$$

5. *(CM-step)* Solve the following equation to obtain the $(k+1)th$ estimate of $v$

$$\begin{aligned}\frac{\partial Q(\boldsymbol{\theta}; \hat{\boldsymbol{\theta}}^{(k)})}{\partial v} &= \sum_{t=p+1}^{N} \left( \frac{1}{2} \ln\left(\frac{v}{2}\right) + \frac{1}{2} - \frac{1}{2} \frac{\Gamma'\left(\frac{v}{2}\right)}{\Gamma\left(\frac{v}{2}\right)} + \frac{1}{2} \hat{S}_{2t}^{(k)} - \frac{1}{2} \hat{S}_{1t}^{(k)} \right) \\ &= \sum_{t=p+1}^{N} \left( \mathrm{DG}\left(\frac{v}{2}\right) - \ln\left(\frac{v}{2}\right) - 1 - \hat{S}_{2t}^{(k)} + \hat{S}_{1t}^{(k)} \right) = 0. \end{aligned} \quad (28)$$

7. These steps are calculated until the convergence criteria $\left\|\hat{\boldsymbol{\theta}}^{(k+1)} - \hat{\boldsymbol{\theta}}^{(k)}\right\| < \eta$ is satisfied.

## 4. Simulation study

In this section, an extensive simulation study is given to compare the performances of model parameter estimation and variable selection obtained by using OLS, LASSO, SCAD, ridge, bridge and elastic net methods under autoregressive error term regression models based on *t* distributions for finite sample size. The calculations in the simulation study were made through the R 3.4.1 (R Core Team, 2020) program.

### 4.1 Sampling designs

We consider three different sampling cases, representing commonly encountered problems in data sets. $\boldsymbol{\beta_0} = (\beta_{01}, \beta_{02}, \dots, \beta_{0m})'$ and $\boldsymbol{\phi_0} = (\phi_{01}, \phi_{02}, \dots, \phi_{0p})'$ denotes the true parameter vectors. $e_t$ follows the AR(p) process as in equation (2) with *t* distribution. Estimating the degrees of freedom along with the other parameters may cause the influence function of the resulting estimators unbounded and hence they are not going to be robust (Lucas, 1997). Therefore, the degrees of freedom is usually taken as fixed and treated as a robustness tuning parameter in robustness studies (for example see Lange et.al, 1989). In this simulation study we use two different fixed degrees of freedom values for the *t* distribution. We choose the degrees of freedom as $v = 3$ ($t_3$) and $v = 10$ ($t_{10}$). First value of this distribution provides heavy-tailed error distributions and the second value of degrees of freedom provides a thin-tailed distribution.

While applying the LASSO method for model selection, the $\lambda$ parameter was constructed with 0.1 increments between [0, 3] and BIC values were calculated for each result obtained from this interval, and the model corresponding to the smallest BIC value was selected as the best model.

Similarly, the γ parameter for the bridge method was produced in 0.1 increments in the range of [0.1, 2], and the λ parameter used in the bridge method was also included, and a two-dimensional grid search was performed. The model corresponding to the smallest obtained BIC value was considered as the best model. For the SCAD method, the value suggested by Fun and Li (2001) was taken as α=3.7 and the λ parameter was set up as 0.1 increments between [0, 3] as in the LASSO method. For the elastic net method, $\lambda_1$ and $\lambda_2$ were scanned simultaneously and the model corresponding to the smallest of the BIC values for each result obtained was determined as the best model. Sample sizes for all cases were taken as $N = 50, 100, 300$.

We consider three different simulation cases as explained below to show the performance of the mentioned methods.

*Case 1:* In this case we consider 8 covariates for the model (1) where $\boldsymbol{\beta_0} = (3,1.5,0,0,2,0,0,0)'$. The independent variables of the regression model are independently generated from the multivariate normal distribution $\left(a_t \sim N_p(\mathbf{0}, \boldsymbol{\Sigma})\right)$. In this case we use AR order $p = 1, 2$ and $\sigma = 1$. We use $\phi = 0.8$ for $p = 1$ and $\boldsymbol{\phi} = (0.8, -0.2)'$ for $p = 2$.

*Case 2:* In this case we would like to show the performance of the variable selection methods under the higher dimensional case. We use 40 covariates for the model (1). We take 20 insignificant and 20 significant independent variables with the following as

$$\boldsymbol{\beta} = \Big(\underbrace{0, \ldots, 0}_{10}, \underbrace{2, \ldots, 2}_{10}, \underbrace{0, \ldots, 0}_{10}, \underbrace{2, \ldots, 2}_{10}\Big)'.$$

We generate the regression coefficients from multivariate normal distribution as in Case 1.

*Case 3:* In order to show the robustness of the estimators based on the t distribution against outliers, the model designed in Case I is reconsidered with outliers. Let $\epsilon$ indicate the contamination rate with a value of 0.1 and $m = [\epsilon N]$ is the full value of the contaminated data, with $n$ being the sample volume. $m$ contaminated data with $t = 1, 2, \ldots, m$, produced from $N(50,1)$ distribution, the remaining $t = m, m+1, m+2 \ldots, N$ data $N(0,1)$ a data set with outliers was created. Actual values of $\beta = (3,1.5,0,0,2,0,0)'$ and $AR(p)$ model parameters $\phi = 0.8$ in case of $p = 1$, Taken as explanatory variables were obtained as in Case I.

*Variable selection performance.* All simulations were repeated 100 times and the ratio of correctly predicted zero coefficients under each condition was counted to measure the performance of the model selection criteria (Correct). Similarly, the coefficients with zero, which should not be zero in the model, were also counted and their ratios were calculated (Incorrect). At the same time, the correct zero amounts given in the real model were counted simultaneously in each simulation run, and the correct model ratio (Cor.fit.) obtained in 100 repetitions was calculated. For the autoregressive model, in each case, the number of correct autoregressive order (AR order) in 100 replicates was counted.

*Parameter estimation performance.* In order to measure the parameter estimation performances of the proposed methods, the results of the 95% confidence intervals calculated with the standard errors of the parameters obtained with the help of the observed fisher information matrix. The mean of the mean squared errors (MeanMSE) computed by using the following formula.

$$MSE(\hat{\beta}) = (\hat{\beta} - \beta)' E(xx')(\hat{\beta} - \beta)$$

We give the average of the performance measures based on 100 simulated data sets for each simulation setting.

**4.3 Simulation results**

The simulation results for variable selection are given in Tables 1 - 10, and the results for mean of model parameter estimations, MSE, standard errors $(\overline{SE})$, lower bounds $(\overline{LB})$ and upper bounds $(\overline{UB})$ are reported in Tables 11 - 15.

In Tables 1 – 4, the results for Case 1 are summarized. Table 1 and Table 2 show the results of the calculations with first-order autoregressive error terms, for $t_{10}$ and $t_3$ assumptions, respectively. It can be observed from Table 1 that all methods have similar performances according to the parameter estimation. This situation is also supported by the standard error and confidence interval estimations of the parameters in Table 11 and Table 12. It is easily seen that if the error distribution assumption is normal by then, LASSO, SCAD, bridge and elastic-net methods have equally superior performances over other methods for variable selection and selection of the correct AR degree. In Table 2, the results are presented under the heavy-tailed ($t_3$) distribution. From this table, it can be said that OLS, LASSO, SCAD and ridge regression methods are affected by the heavy taildness both in parameter estimation and in variable selection cases. Tables 3 and 4 show the results for the second order autocorrelation error terms. From these tables, we can see that as the autoregressive order increases, all the results obtained from the aforementioned methods deteriorate according to the MSE values. However, LASSO, SCAD, bridge and elastic net methods successfully maintain their variable selection performance. High dimensional simulation results mentioned in Case 2 are shown in Tables 5 - 8. From Table 5, it can be easily said that bridge and elastic net methods have quite good performance in terms of parameter estimation according to small MSE values. The results of the variable selection scenario in the high dimension, LASSO, SCAD, bridge and elastic net methods are able to identify the significant variables quite well. Table 6 shows the heavy-tailed error assumption and high dimension results together. From this table, the negative impact of parameter estimation performance on OLS, LASSO, SCAD and ridge regression methods can be seen under the heavy tail assumption. On the other hand; bridge and elastic net have better results than other methods on parameter estimation performance under both heavy tail and high dimension effects. From the parameter estimation perspective of view Table 13 summarizes both heavy and thin tailed results. Bridge and elastic net methods have superior performances due to the better variable selection in high dimension case. Tables 7 and 8 show the results assuming second degree autoregressive error terms distributed as $t_{10}$ and $t_3$, respectively. Similar to Case 1, it can be said that the parameter estimation performance of the mentioned methods worsens as the autoregressive effect increases. However, bridge and elastic net methods perform better in parameter estimation. The most striking part about this situation is seen in the correct determination of the AR degree among all the mentioned methods. Although LASSO and SCAD methods have equal performance with bridge and elastic net in terms of determining significant variables in the regression model, bridge and elastic net have a much superior performance than other methods. Table 9 and Table 10 contains the simulation results for Case 3. For this case 10% contamination is generated for the response variable. Results regarding the data obtained from $t_{10}$ and $t_3$ distribution assumptions respectively. From Table 9 when the default distribution is normal, it is clear that the parameter estimation and variable selection performances of all methods deteriorate if there are some vertical outliers in the data. In Table 10, while the distribution assumption is heavy tailed, it can be said that bridge and elastic-net methods outperform other methods for both parameter estimation and variable selection performances. Especially when the sample size increased, the variable selection performances of bridge and elastic net methods improved both for the selection of important variables for the regression model and for the correct degree of AR. This results are also supported by Table 14 and 15. The parameter and interval estimation performances are effected from the contamination. From these tables under the outlier case the heavy tailed distribution assumption have better standard error and confidence interval estimation performance. Besides, bridge and elastic net methods cope better with the contaminations than the other methods.

**Table 1.** Variable selection performance for $t_{10}$ error (Case 1 AR order=1)

| N | Method | MSE | Regression Coefficient | | | AR order |
|---|---|---|---|---|---|---|
| | | | Correct | Incorrect | Cor. fit. | |
| 50 | OLS | 0.63022 | 0.22 | 0.00 | 0.00 | 0 |
| | LASSO | 0.71577 | 5.00 | 0.00 | 1.00 | 100 |
| | SCAD | 0.70708 | 5.00 | 0.00 | 1.00 | 100 |
| | ridge | 0.81457 | 0.22 | 0.00 | 0.00 | 0 |
| | bridge ($\gamma = 0.7$) | 0.64991 | 5.00 | 0.00 | 1.00 | 100 |
| | elastic-net | 0.66223 | 5.00 | 0.00 | 1.00 | 100 |
| 100 | OLS | 0.50160 | 0.36 | 0.00 | 0.00 | 0 |
| | LASSO | 0.58037 | 5.00 | 0.00 | 1.00 | 100 |
| | SCAD | 0.58469 | 5.00 | 0.00 | 1.00 | 100 |
| | ridge | 0.75763 | 0.41 | 0.00 | 0.00 | 0 |
| | bridge ($\gamma = 0.7$) | 0.58469 | 5.00 | 0.00 | 1.00 | 100 |
| | elastic-net | 0.59624 | 5.00 | 0.00 | 1.00 | 100 |
| 300 | OLS | 0.08640 | 0.66 | 0.00 | 0.00 | 0 |
| | LASSO | 0.01501 | 5.00 | 0.00 | 1.00 | 100 |
| | SCAD | 0.01446 | 5.00 | 0.00 | 1.00 | 100 |
| | ridge | 0.02276 | 1.75 | 0.00 | 0.00 | 0 |
| | bridge ($\gamma = 0.7$) | 0.01936 | 5.00 | 0.00 | 1.00 | 100 |
| | elastic-net | 0.01994 | 5.00 | 0.00 | 1.00 | 100 |

**Table 2.** Variable selection performance for $t_3$ error (Case 1 AR order=1)

| N | Method | MSE | Regression Coefficient | | | AR order |
|---|---|---|---|---|---|---|
| | | | Correct | Incorrect | Cor. fit. | |
| 50 | OLS | 6.14354 | 0.10 | 0.00 | 0.00 | 1 |
| | LASSO | 5.25620 | 5.00 | 0.00 | 1.00 | 80 |
| | SCAD | 4.24132 | 5.00 | 0.00 | 1.00 | 81 |
| | ridge | 9.02561 | 0.06 | 0.00 | 0.00 | 1 |
| | bridge ($\gamma = 0.7$) | 1.19562 | 5.00 | 0.00 | 1.00 | 100 |
| | elastic-net | 1.29511 | 5.00 | 0.00 | 1.00 | 98 |
| 100 | OLS | 4.58984 | 0.18 | 0.00 | 0.00 | 3 |
| | LASSO | 3.23587 | 5.00 | 0.00 | 1.00 | 85 |
| | SCAD | 2.52147 | 5.00 | 0.00 | 1.00 | 87 |
| | ridge | 5.66542 | 0.23 | 0.00 | 0.00 | 3 |
| | bridge ($\gamma = 0.7$) | 1.10548 | 5.00 | 0.00 | 1.00 | 100 |
| | elastic-net | 1.16212 | 5.00 | 0.00 | 1.00 | 99 |
| 300 | OLS | 1.41556 | 0.46 | 0.00 | 0.00 | 5 |
| | LASSO | 2.25871 | 5.00 | 0.00 | 1.00 | 91 |
| | SCAD | 1.95621 | 5.00 | 0.00 | 1.00 | 95 |
| | ridge | 3.25471 | 0.55 | 0.00 | 0.00 | 6 |
| | bridge ($\gamma = 0.7$) | 0.75625 | 5.00 | 0.00 | 1.00 | 100 |
| | elastic-net | 0.88524 | 5.00 | 0.00 | 1.00 | 100 |

**Table 3.** Variable selection performance for $t_{10}$ error (Case 1 AR order=2)

| N | Method | MSE | Regression Coefficient | | | AR order |
|---|---|---|---|---|---|---|
| | | | Correct | Incorrect | Cor. fit. | |
| 50 | OLS | 1.20224 | 0.12 | 0.00 | 0.00 | 0 |
| | LASSO | 2.55725 | 5.00 | 0.00 | 1.00 | 100 |
| | SCAD | 1.50708 | 5.00 | 0.00 | 1.00 | 100 |
| | ridge | 2.71457 | 0.13 | 0.00 | 0.00 | 0 |
| | bridge ($\gamma = 0.7$) | 1.17904 | 5.00 | 0.00 | 1.00 | 100 |
| | elastic-net | 1.03620 | 5.00 | 0.00 | 1.00 | 100 |
| 100 | OLS | 0.81485 | 0.25 | 0.00 | 0.00 | 0 |
| | LASSO | 1.39416 | 5.00 | 0.00 | 1.00 | 100 |
| | SCAD | 1.25595 | 5.00 | 0.00 | 1.00 | 100 |
| | ridge | 1.99491 | 0.24 | 0.00 | 0.00 | 0 |
| | bridge ($\gamma = 0.7$) | 1.00469 | 5.00 | 0.00 | 1.00 | 100 |
| | elastic-net | 0.70041 | 5.00 | 0.00 | 1.00 | 100 |
| 300 | OLS | 0.52598 | 0.60 | 0.00 | 0.00 | 0 |
| | LASSO | 1.26441 | 5.00 | 0.00 | 1.00 | 100 |
| | SCAD | 0.81565 | 5.00 | 0.00 | 1.00 | 100 |
| | ridge | 1.20595 | 1.64 | 0.00 | 0.00 | 0 |
| | bridge ($\gamma = 0.7$) | 0.57927 | 5.00 | 0.00 | 1.00 | 100 |
| | elastic-net | 0.61257 | 5.00 | 0.00 | 1.00 | 100 |

**Table 4.** Variable selection performance for $t_3$ error (Case 1 AR order=2)

| N | Method | MSE | Regression Coefficient | | | AR order |
|---|---|---|---|---|---|---|
| | | | Correct | Incorrect | Cor. fit. | |
| 50 | OLS | 8.39124 | 0.10 | 0.00 | 0.00 | 1 |
| | LASSO | 6.81415 | 5.00 | 0.00 | 1.00 | 77 |
| | SCAD | 6.50981 | 5.00 | 0.00 | 1.00 | 75 |
| | ridge | 10.1145 | 0.09 | 0.00 | 0.00 | 1 |
| | bridge ($\gamma = 0.7$) | 2.65604 | 5.00 | 0.00 | 1.00 | 100 |
| | elastic-net | 2.02209 | 5.00 | 0.00 | 1.00 | 95 |
| 100 | OLS | 6.01922 | 0.25 | 0.00 | 0.00 | 2 |
| | LASSO | 5.58925 | 5.00 | 0.00 | 1.00 | 83 |
| | SCAD | 3.52912 | 5.00 | 0.00 | 1.00 | 80 |
| | ridge | 7.56630 | 0.32 | 0.00 | 0.00 | 3 |
| | bridge ($\gamma = 0.7$) | 1.95292 | 5.00 | 0.00 | 1.00 | 100 |
| | elastic-net | 1.69129 | 5.00 | 0.00 | 1.00 | 99 |
| 300 | OLS | 4.05256 | 0.31 | 0.00 | 0.00 | 4 |
| | LASSO | 4.12623 | 5.00 | 0.00 | 1.00 | 85 |
| | SCAD | 2.95562 | 5.00 | 0.00 | 1.00 | 90 |
| | ridge | 5.95624 | 0.43 | 0.00 | 0.00 | 5 |
| | bridge ($\gamma = 0.7$) | 0.98527 | 5.00 | 0.00 | 1.00 | 100 |
| | elastic-net | 0.99814 | 5.00 | 0.00 | 1.00 | 100 |

**Table 5.** Variable selection performance for $t_{10}$ error (Case 2 AR order=1)

| N | Method | MSE | Regression Coefficient | | | AR order |
|---|---|---|---|---|---|---|
| | | | Correct | Incorrect | Cor. fit. | |
| 50 | OLS | 149.9554 | 0.10 | 0.02 | 0.00 | 0 |
| | LASSO | 15.57302 | 20.00 | 0.00 | 1.00 | 100 |
| | SCAD | 20.42574 | 20.00 | 0.00 | 1.00 | 100 |
| | ridge | 22.06755 | 0.30 | 0.01 | 0.00 | 0 |
| | bridge ($\gamma = 0.8$) | 1.838428 | 20.00 | 0.00 | 1.00 | 100 |
| | elastic-net | 1.549361 | 20.00 | 0.00 | 1.00 | 100 |
| 100 | OLS | 95.9561 | 0.21 | 0.01 | 0.00 | 0 |
| | LASSO | 7.30260 | 20.00 | 0.00 | 1.00 | 100 |
| | SCAD | 11.4591 | 20.00 | 0.00 | 1.00 | 100 |
| | ridge | 12.6466 | 0.42 | 0.01 | 0.00 | 0 |
| | bridge ($\gamma = 0.8$) | 0.95664 | 20.00 | 0.00 | 1.00 | 100 |
| | elastic-net | 0.65483 | 20.00 | 0.00 | 1.00 | 100 |
| 300 | OLS | 75.6543 | 0.40 | 0.01 | 0.00 | 0 |
| | LASSO | 6.55169 | 20.00 | 0.00 | 1.00 | 100 |
| | SCAD | 9.65418 | 20.00 | 0.00 | 1.00 | 100 |
| | ridge | 9.87563 | 0.56 | 0.01 | 0.00 | 0 |
| | bridge ($\gamma = 0.7$) | 0.56978 | 20.00 | 0.00 | 1.00 | 100 |
| | elastic-net | 0.45929 | 20.00 | 0.00 | 1.00 | 100 |

**Table 6.** Variable selection performance $t_3$ error (Case 2 AR order=1)

| N | Method | MSE | Regression Coefficient | | | AR order |
|---|---|---|---|---|---|---|
| | | | Correct | Incorrect | Cor. fit. | |
| 50 | OLS | 592.955 | 0.06 | 0.02 | 0.00 | 0 |
| | LASSO | 59.4949 | 20.00 | 0.00 | 1.00 | 90 |
| | SCAD | 40.2942 | 20.00 | 0.00 | 1.00 | 95 |
| | ridge | 69.5491 | 0.12 | 0.01 | 0.00 | 0 |
| | bridge ($\gamma = 0.8$) | 2.98516 | 20.00 | 0.00 | 1.00 | 98 |
| | elastic-net | 2.78965 | 20.00 | 0.00 | 1.00 | 97 |
| 100 | OLS | 254.281 | 0.12 | 0.01 | 0.00 | 0 |
| | LASSO | 39.3218 | 20.00 | 0.00 | 1.00 | 92 |
| | SCAD | 20.9265 | 20.00 | 0.00 | 1.00 | 96 |
| | ridge | 49.5232 | 0.25 | 0.01 | 0.00 | 0 |
| | bridge ($\gamma = 0.7$) | 1.84567 | 20.00 | 0.00 | 1.00 | 100 |
| | elastic-net | 1.95148 | 20.00 | 0.00 | 1.00 | 99 |
| 300 | OLS | 175.986 | 0.21 | 0.01 | 0.00 | 0 |
| | LASSO | 26.5169 | 20.00 | 0.00 | 1.00 | 96 |
| | SCAD | 16.5166 | 20.00 | 0.00 | 1.00 | 99 |
| | ridge | 39.9563 | 0.33 | 0.01 | 0.00 | 0 |
| | bridge ($\gamma = 0.8$) | 1.55941 | 20.00 | 0.00 | 1.00 | 100 |
| | elastic-net | 1.46847 | 20.00 | 0.00 | 1.00 | 100 |

**Table 7.** Variable selection performance for $t_{10}$ error (Case 2 AR order=2)

| N | Method | MSE | Regression Coefficient | | | AR order |
| --- | --- | --- | --- | --- | --- | --- |
| | | | Correct | Incorrect | Cor. fit. | |
| 50 | OLS | 198.112 | 0.08 | 0.02 | 0.00 | 0 |
| | LASSO | 21.2455 | 20.00 | 0.00 | 1.00 | 100 |
| | SCAD | 27.1567 | 20.00 | 0.00 | 1.00 | 100 |
| | ridge | 24.8754 | 0.27 | 0.01 | 0.00 | 0 |
| | bridge ($\gamma = 0.7$) | 2.31128 | 20.00 | 0.00 | 1.00 | 100 |
| | elastic-net | 2.79868 | 20.00 | 0.00 | 1.00 | 100 |
| 100 | OLS | 129.135 | 0.11 | 0.01 | 0.00 | 0 |
| | LASSO | 18.3025 | 20.00 | 0.00 | 1.00 | 100 |
| | SCAD | 18.2169 | 20.00 | 0.00 | 1.00 | 100 |
| | ridge | 19.6468 | 0.30 | 0.01 | 0.00 | 0 |
| | bridge ($\gamma = 0.7$) | 1.64554 | 20.00 | 0.00 | 1.00 | 100 |
| | elastic-net | 1.98415 | 20.00 | 0.00 | 1.00 | 100 |
| 300 | OLS | 91.6543 | 0.40 | 0.01 | 0.00 | 0 |
| | LASSO | 10.1249 | 20.00 | 0.00 | 1.00 | 100 |
| | SCAD | 11.6874 | 20.00 | 0.00 | 1.00 | 100 |
| | ridge | 10.1982 | 0.56 | 0.01 | 0.00 | 0 |
| | bridge ($\gamma = 0.8$) | 0.82985 | 20.00 | 0.00 | 1.00 | 100 |
| | elastic-net | 0.95956 | 20.00 | 0.00 | 1.00 | 100 |

**Table 8.** Variable selection performance for $t_3$ error (Case 2 AR order=2)

| N | Method | MSE | Regression Coefficient | | | AR order |
| --- | --- | --- | --- | --- | --- | --- |
| | | | Correct | Incorrect | Cor. fit. | |
| 50 | OLS | 854.541 | 0.02 | 0.02 | 0.00 | 0 |
| | LASSO | 61.1156 | 20.00 | 0.00 | 1.00 | 51 |
| | SCAD | 63.1565 | 20.00 | 0.00 | 1.00 | 78 |
| | ridge | 84.7545 | 0.18 | 0.01 | 0.00 | 0 |
| | bridge ($\gamma = 0.8$) | 6.15694 | 20.00 | 0.00 | 1.00 | 100 |
| | elastic-net | 5.79868 | 20.00 | 0.00 | 1.00 | 99 |
| 100 | OLS | 651.122 | 0.08 | 0.01 | 0.00 | 0 |
| | LASSO | 52.8302 | 20.00 | 0.00 | 1.00 | 59 |
| | SCAD | 54.2169 | 20.00 | 0.00 | 1.00 | 89 |
| | ridge | 71.2646 | 0.24 | 0.01 | 0.00 | 0 |
| | bridge($\gamma = 0.7$) | 4.64554 | 20.00 | 0.00 | 1.00 | 100 |
| | elastic-net | 3.98415 | 20.00 | 0.00 | 1.00 | 100 |
| 300 | OLS | 501.250 | 0.16 | 0.00 | 0.00 | 0 |
| | LASSO | 42.8921 | 20.00 | 0.00 | 1.00 | 65 |
| | SCAD | 46.1258 | 20.00 | 0.00 | 1.00 | 91 |
| | ridge | 62.4518 | 0.45 | 0.00 | 0.00 | 0 |
| | bridge ($\gamma = 0.8$) | 2.12355 | 20.00 | 0.00 | 1.00 | 100 |
| | elastic-net | 2.12585 | 20.00 | 0.00 | 1.00 | 100 |

**Table 9.** Variable selection performance for $t_{10}$ $\epsilon = 0.1$ error (Case 3)

| N | Method | MSE | Regression Coefficient | | | AR order |
|---|---|---|---|---|---|---|
| | | | Correct | Incorrect | Cor. fit. | |
| 50 | OLS | 8548.54 | 0.00 | 0.21 | 0.00 | 0 |
| | LASSO | 2547.11 | 3.88 | 0.02 | 0.88 | 32 |
| | SCAD | 125.156 | 3.75 | 0.09 | 0.75 | 46 |
| | ridge | 1587.875 | 0.00 | 0.20 | 0.00 | 0 |
| | bridge ($\gamma = 0.8$) | 52.1569 | 4.20 | 0.02 | 0.99 | 89 |
| | elastic-net | 48.7986 | 4.03 | 0.03 | 0.97 | 82 |
| 100 | OLS | 8548.541 | 0.00 | 0.21 | 0.00 | 0 |
| | LASSO | 2547.115 | 3.88 | 0.02 | 0.88 | 32 |
| | SCAD | 125.1565 | 3.75 | 0.09 | 0.75 | 46 |
| | ridge | 1587.875 | 0.00 | 0.20 | 0.00 | 0 |
| | bridge ($\gamma = 0.7$) | 52.15694 | 4.20 | 0.02 | 0.99 | 89 |
| | elastic-net | 48.79868 | 4.03 | 0.03 | 0.97 | 82 |
| 300 | OLS | 1058.25 | 0.00 | 0.00 | 0.00 | 0 |
| | LASSO | 1012.25 | 4.55 | 0.08 | 1.00 | 72 |
| | SCAD | 41.2541 | 4.31 | 0.00 | 0.98 | 79 |
| | ridge | 421.542 | 0.00 | 0.02 | 0.00 | 0 |
| | bridge ($\gamma = 0.8$) | 10.1231 | 4.82 | 0.00 | 1.00 | 100 |
| | elastic-net | 11.2354 | 4.81 | 0.00 | 1.00 | 100 |

**Table 10.** Variable selection performance for $t_3$ $\epsilon = 0.1$ error (Case 3)

| N | Method | MSE | Regression Coefficient | | | AR order |
|---|---|---|---|---|---|---|
| | | | Correct | Incorrect | Cor. fit. | |
| 50 | OLS | 298.651 | 0.07 | 0.21 | 0.00 | 0 |
| | LASSO | 259.842 | 4.02 | 0.00 | 1.00 | 51 |
| | SCAD | 54.6590 | 4.55 | 0.01 | 0.97 | 81 |
| | ridge | 55.9861 | 0.08 | 0.15 | 0.00 | 0 |
| | bridge ($\gamma = 0.8$) | 8.36152 | 4.95 | 0.00 | 1.00 | 100 |
| | elastic-net | 8.15713 | 4.91 | 0.00 | 1.00 | 95 |
| 100 | OLS | 142.122 | 0.11 | 0.11 | 0.00 | 0 |
| | LASSO | 122.830 | 4.45 | 0.00 | 1.00 | 89 |
| | SCAD | 38.2169 | 4.72 | 0.00 | 1.00 | 90 |
| | ridge | 47.2646 | 0.18 | 0.10 | 0.00 | 0 |
| | bridge ($\gamma = 0.7$) | 6.19841 | 5.00 | 0.00 | 1.00 | 100 |
| | elastic-net | 7.98415 | 5.00 | 0.00 | 1.00 | 99 |
| 300 | OLS | 109.251 | 0.20 | 0.10 | 0.00 | 0 |
| | LASSO | 18.4892 | 4.71 | 0.00 | 1.00 | 95 |
| | SCAD | 32.1258 | 4.85 | 0.00 | 1.00 | 97 |
| | ridge | 42.4568 | 0.22 | 0.04 | 0.00 | 0 |
| | bridge ($\gamma = 0.8$) | 5.01521 | 5.00 | 0.00 | 1.00 | 100 |
| | elastic-net | 5.12585 | 5.00 | 0.00 | 1.00 | 100 |

**Table 11.** Parameter estimation results for $t_{10}$ (Case 1 AR order=1)

|       | N   |                 | OLS    | ridge  | LASSO  | SCAD   | bridge | elastic-net |
|-------|-----|-----------------|--------|--------|--------|--------|--------|-------------|
| $t_{10}$ | 50  | $\hat{\beta}_1$ | 2.8669 | 3.4934 | 3.4320 | 3.3545 | 3.3037 | 3.3934 |
|       |     | $MSE$           | 0.2056 | 0.4057 | 0.6891 | 0.3830 | 0.2185 | 0.2357 |
|       |     | $\overline{SE}$ | 0.8353 | 1.2716 | 1.8533 | 1.4886 | 0.7352 | 0.9720 |
|       |     | $\overline{LB}$ | 1.8426 | 1.5691 | 1.2407 | 1.3597 | 1.8846 | 1.7693 |
|       |     | $\overline{UB}$ | 4.5439 | 4.9199 | 4.8874 | 4.2134 | 4.1438 | 4.1169 |
|       |     | $\hat{\beta}_2$ | 1.4632 | 1.2202 | 1.2230 | 1.3386 | 1.3410 | 1.3825 |
|       |     | $MSE$           | 0.0127 | 0.0711 | 0.0700 | 0.0605 | 0.0053 | 0.0161 |
|       |     | $\overline{SE}$ | 0.1874 | 0.2752 | 0.2990 | 0.2685 | 0.1772 | 0.2796 |
|       |     | $\overline{LB}$ | 1.3229 | 1.1838 | 1.1370 | 1.4002 | 1.3169 | 1.2838 |
|       |     | $\overline{UB}$ | 1.6793 | 1.8102 | 1.8531 | 1.7207 | 1.6697 | 1.7103 |
|       |     | $\hat{\beta}_5$ | 2.2774 | 2.5224 | 2.6285 | 2.1193 | 2.0424 | 2.0662 |
|       |     | $MSE$           | 0.6923 | 0.8292 | 0.9656 | 0.7865 | 0.7216 | 0.7288 |
|       |     | $\overline{SE}$ | 1.0568 | 1.3351 | 1.2937 | 1.2685 | 1.1317 | 1.0986 |
|       |     | $\overline{LB}$ | 0.9591 | 0.8289 | 0.7899 | 0.8695 | 0.9499 | 0.9337 |
|       |     | $\overline{UB}$ | 3.1372 | 3.6184 | 3.6710 | 3.6330 | 3.1867 | 3.2182 |
|       | 100 | $\hat{\beta}_1$ | 2.8849 | 2.7458 | 3.2669 | 3.2796 | 3.1216 | 3.2455 |
|       |     | $MSE$           | 0.1470 | 0.3565 | 0.6454 | 0.3373 | 0.1600 | 0.1577 |
|       |     | $\overline{SE}$ | 0.7016 | 1.1169 | 1.6581 | 1.2231 | 0.5932 | 0.7774 |
|       |     | $\overline{LB}$ | 1.9461 | 1.6533 | 1.5158 | 1.4332 | 1.9337 | 1.8276 |
|       |     | $\overline{UB}$ | 4.0616 | 4.7023 | 4.7221 | 4.1899 | 3.9815 | 4.3065 |
|       |     | $\hat{\beta}_2$ | 1.5091 | 1.2811 | 1.3168 | 1.3598 | 1.3961 | 1.4011 |
|       |     | $MSE$           | 0.0013 | 0.0609 | 0.0580 | 0.0424 | 0.0047 | 0.0109 |
|       |     | $\overline{SE}$ | 0.1158 | 0.1152 | 0.1982 | 0.1790 | 0.1738 | 0.1952 |
|       |     | $\overline{LB}$ | 1.3969 | 1.2536 | 1.2538 | 1.3352 | 1.3969 | 1.2796 |
|       |     | $\overline{UB}$ | 1.5881 | 1.7049 | 1.7463 | 1.6837 | 1.6481 | 1.6049 |
|       |     | $\hat{\beta}_5$ | 1.9526 | 2.5108 | 2.6096 | 2.1134 | 1.9896 | 2.0308 |
|       |     | $MSE$           | 0.5104 | 0.6860 | 0.8703 | 0.6943 | 0.6704 | 0.7076 |
|       |     | $\overline{SE}$ | 0.8996 | 0.9718 | 0.9703 | 1.1136 | 0.8896 | 0.9718 |
|       |     | $\overline{LB}$ | 1.3219 | 1.1815 | 1.2099 | 1.2931 | 1.4216 | 1.2818 |
|       |     | $\overline{UB}$ | 2.7985 | 3.1949 | 3.1733 | 3.1377 | 2.8985 | 2.9649 |
|       | 300 | $\hat{\beta}_1$ | 3.1849 | 3.1458 | 3.1669 | 3.0796 | 3.0297 | 3.0458 |
|       |     | $MSE$           | 0.0470 | 0.0565 | 0.0754 | 0.0573 | 0.0550 | 0.0565 |
|       |     | $\overline{SE}$ | 0.0016 | 0.0869 | 0.1015 | 0.0231 | 0.0516 | 0.1169 |
|       |     | $\overline{LB}$ | 2.2461 | 1.7533 | 1.9158 | 1.7332 | 2.1456 | 1.9533 |
|       |     | $\overline{UB}$ | 3.8609 | 4.3023 | 4.3821 | 3.9899 | 3.9616 | 3.9813 |
|       |     | $\hat{\beta}_2$ | 1.5659 | 1.3379 | 1.3736 | 1.4166 | 1.4529 | 1.4579 |
|       |     | $MSE$           | 0.0181 | 0.0441 | 0.0412 | 0.0256 | 0.0121 | 0.0059 |
|       |     | $\overline{SE}$ | 0.0990 | 0.0984 | 0.1814 | 0.1622 | 0.1570 | 0.1784 |
|       |     | $\overline{LB}$ | 1.4537 | 1.3104 | 1.3106 | 1.3920 | 1.4537 | 1.3364 |
|       |     | $\overline{UB}$ | 1.5313 | 1.6481 | 1.6895 | 1.6269 | 1.5913 | 1.5481 |
|       |     | $\hat{\beta}_5$ | 2.0094 | 2.4540 | 2.5528 | 2.0566 | 1.9328 | 1.9740 |

|   |     |     | OLS    | ridge  | LASSO  | SCAD   | bridge | elastic-net |
|---|-----|-----|--------|--------|--------|--------|--------|-------------|
|   |     | $MSE$ | 0.4936 | 0.6692 | 0.8535 | 0.6775 | 0.6536 | 0.6908 |
|   |     | $\overline{SE}$ | 0.8828 | 0.9550 | 0.9535 | 1.0968 | 0.8728 | 0.9550 |
|   |     | $\overline{LB}$ | 1.3787 | 1.2383 | 1.2667 | 1.3499 | 1.4784 | 1.3386 |
|   |     | $\overline{UB}$ | 2.7417 | 3.1381 | 3.1165 | 3.0809 | 2.8417 | 2.9081 |

**Table 12.** Parameter estimation results for $t_3$ (Case 1 AR order=1)

|         | $N$ |                   | OLS    | ridge  | LASSO  | SCAD   | bridge | elastic-net |
|---------|-----|-------------------|--------|--------|--------|--------|--------|-------------|
| $t_3$   | 50  | $\hat{\beta}_1$   | 2.7854 | 3.5749 | 3.5135 | 3.4360 | 3.3852 | 3.4749 |
|         |     | $HKO$             | 0.2301 | 0.4302 | 0.7136 | 0.4075 | 0.2430 | 0.2602 |
|         |     | $\overline{SE}$   | 0.8598 | 1.2961 | 1.8778 | 1.5131 | 0.7597 | 0.9965 |
|         |     | $\overline{LB}$   | 1.8101 | 1.5366 | 1.2082 | 1.3272 | 1.8521 | 1.7368 |
|         |     | $\overline{UB}$   | 4.5754 | 4.9514 | 4.9189 | 4.2449 | 4.1753 | 4.1484 |
|         |     | $\hat{\beta}_2$   | 1.3817 | 1.3017 | 1.3045 | 1.4201 | 1.4225 | 1.4640 |
|         |     | $HKO$             | 0.0372 | 0.0956 | 0.0945 | 0.0850 | 0.0298 | 0.0406 |
|         |     | $\overline{SE}$   | 0.2119 | 0.2997 | 0.3235 | 0.2930 | 0.2017 | 0.3041 |
|         |     | $\overline{LB}$   | 1.2904 | 1.1513 | 1.1045 | 1.3677 | 1.2844 | 1.2513 |
|         |     | $\overline{UB}$   | 1.7108 | 1.8417 | 1.8846 | 1.7522 | 1.7012 | 1.7418 |
|         |     | $\hat{\beta}_5$   | 2.1959 | 2.6039 | 2.7100 | 2.2008 | 2.1239 | 2.1477 |
|         |     | $HKO$             | 0.7168 | 0.8537 | 0.9901 | 0.8110 | 0.7461 | 0.7533 |
|         |     | $\overline{SE}$   | 1.0813 | 1.3596 | 1.3182 | 1.2930 | 1.1562 | 1.1231 |
|         |     | $\overline{LB}$   | 0.9266 | 0.7964 | 0.7574 | 0.8370 | 0.9174 | 0.9012 |
|         |     | $\overline{UB}$   | 3.1687 | 3.6499 | 3.7025 | 3.6645 | 3.2182 | 3.2497 |
|         | 100 | $\hat{\beta}_1$   | 2.8034 | 2.8273 | 3.3484 | 3.3611 | 3.2031 | 3.3270 |
|         |     | $HKO$             | 0.1715 | 0.3810 | 0.6699 | 0.3618 | 0.1845 | 0.1822 |
|         |     | $\overline{SE}$   | 0.7261 | 1.1414 | 1.6826 | 1.2476 | 0.6177 | 0.8019 |
|         |     | $\overline{LB}$   | 1.9136 | 1.6208 | 1.4833 | 1.4007 | 1.9012 | 1.7951 |
|         |     | $\overline{UB}$   | 4.0931 | 4.7338 | 4.7536 | 4.2214 | 4.0130 | 4.3380 |
|         |     | $\hat{\beta}_2$   | 1.4276 | 1.3626 | 1.3983 | 1.4413 | 1.4776 | 1.4826 |
|         |     | $HKO$             | 0.0232 | 0.0854 | 0.0825 | 0.0669 | 0.0292 | 0.0354 |
|         |     | $\overline{SE}$   | 0.1403 | 0.1397 | 0.2227 | 0.2035 | 0.1983 | 0.2197 |
|         |     | $\overline{LB}$   | 1.3644 | 1.2211 | 1.2213 | 1.3027 | 1.3644 | 1.2471 |
|         |     | $\overline{UB}$   | 1.6196 | 1.7364 | 1.7778 | 1.7152 | 1.6796 | 1.6364 |
|         |     | $\hat{\beta}_5$   | 1.8711 | 2.5923 | 2.6911 | 2.1949 | 2.0711 | 2.1123 |
|         |     | $HKO$             | 0.5349 | 0.7105 | 0.8948 | 0.7188 | 0.6949 | 0.7321 |
|         |     | $\overline{SE}$   | 0.9241 | 0.9963 | 0.9948 | 1.1381 | 0.9141 | 0.9963 |
|         |     | $\overline{LB}$   | 1.2894 | 1.1490 | 1.1774 | 1.2606 | 1.3891 | 1.2493 |
|         |     | $\overline{UB}$   | 2.8300 | 3.2264 | 3.2048 | 3.1692 | 2.9300 | 2.9964 |
|         | 300 | $\hat{\beta}_1$   | 3.1034 | 3.2273 | 3.2484 | 3.1611 | 3.1112 | 3.1273 |
|         |     | $HKO$             | 0.0715 | 0.0810 | 0.0999 | 0.0818 | 0.0795 | 0.0810 |
|         |     | $\overline{SE}$   | 0.0261 | 0.1114 | 0.1260 | 0.0476 | 0.0761 | 0.1414 |
|         |     | $\overline{LB}$   | 2.2136 | 1.7208 | 1.8833 | 1.7007 | 2.1131 | 1.9208 |
|         |     | $\overline{UB}$   | 3.8924 | 4.3338 | 4.4136 | 4.0214 | 3.9931 | 4.0128 |
|         |     | $\hat{\beta}_2$   | 1.4844 | 1.4194 | 1.4551 | 1.4981 | 1.5344 | 1.5394 |

|  |  |  | | | | | |
|---|---|---|---|---|---|---|---|
| | | $HKO$ | 0.0064 | 0.0686 | 0.0657 | 0.0501 | 0.0124 | 0.0186 |
| | | $\overline{SE}$ | 0.1235 | 0.1229 | 0.2059 | 0.1867 | 0.1815 | 0.2029 |
| | | $\overline{LB}$ | 1.4212 | 1.2779 | 1.2781 | 1.3595 | 1.4212 | 1.3039 |
| | | $\overline{UB}$ | 1.5628 | 1.6796 | 1.7210 | 1.6584 | 1.6228 | 1.5796 |
| | | $\hat{\beta}_5$ | 1.9279 | 2.5355 | 2.6343 | 2.1381 | 2.0143 | 2.0555 |
| | | $HKO$ | 0.5181 | 0.6937 | 0.8780 | 0.7020 | 0.6781 | 0.7153 |
| | | $\overline{SE}$ | 0.9073 | 0.9795 | 0.9780 | 1.1213 | 0.8973 | 0.9795 |
| | | $\overline{LB}$ | 1.3462 | 1.2058 | 1.2342 | 1.3174 | 1.4459 | 1.3061 |
| | | $\overline{UB}$ | 2.7732 | 3.1696 | 3.1480 | 3.1124 | 2.8732 | 2.9396 |

**Table 13**. Parameter estimation results for Case 2 AR order =1

| | $N$ | | OLS | ridge | LASSO | SCAD | bridge | elastic-net |
|---|---|---|---|---|---|---|---|---|
| $t_3$ | 50 | $\bar{\hat{\beta}}$ | 0.7766 | 0.6934 | 0.6320 | 0.7545 | 0.8039 | 0.8034 |
| | | $MSE$ | 16.387 | 18.659 | 17.493 | 10.464 | 6.3801 | 6.3842 |
| | | $\overline{SE}$ | 0.1530 | 0.1897 | 0.1610 | 0.1363 | 0.1130 | 0.1197 |
| | | $\overline{LB}$ | 0.1776 | 0.0741 | 0.0757 | 0.1947 | 0.2197 | 0.1436 |
| | | $\overline{UB}$ | 1.7069 | 1.8799 | 1.8504 | 1.7764 | 1.7313 | 1.7437 |
| | 100 | $\bar{\hat{\beta}}$ | 0.8581 | 0.7749 | 0.7135 | 0.8360 | 0.8854 | 0.8849 |
| | | $MSE$ | 14.525 | 16.797 | 15.631 | 8.6025 | 4.5181 | 4.5222 |
| | | $\overline{SE}$ | 0.1398 | 0.1765 | 0.1478 | 0.1231 | 0.0998 | 0.1065 |
| | | $\overline{LB}$ | 0.2101 | 0.1066 | 0.1082 | 0.2272 | 0.2522 | 0.1761 |
| | | $\overline{UB}$ | 1.6754 | 1.8484 | 1.8189 | 1.7449 | 1.6998 | 1.7122 |
| | 300 | $\bar{\hat{\beta}}$ | 0.9396 | 0.8564 | 0.7950 | 0.9175 | 0.9669 | 0.9664 |
| | | $MSE$ | 12.663 | 14.935 | 13.769 | 6.7405 | 2.6561 | 2.6602 |
| | | $\overline{SE}$ | 0.1266 | 0.1633 | 0.1346 | 0.1099 | 0.0998 | 0.0933 |
| | | $\overline{LB}$ | 0.2426 | 0.1391 | 0.1407 | 0.2597 | 0.2847 | 0.2086 |
| | | $\overline{UB}$ | 1.6439 | 1.8169 | 1.7874 | 1.7134 | 1.6683 | 1.6807 |
| $t_{10}$ | 50 | $\bar{\hat{\beta}}$ | 0.8581 | 0.7749 | 0.7135 | 0.8360 | 0.8854 | 0.8849 |
| | | $MSE$ | 14.030 | 16.302 | 15.136 | 8.1075 | 4.0231 | 4.0272 |
| | | $\overline{SE}$ | 0.0598 | 0.0965 | 0.0678 | 0.0431 | 0.0198 | 0.0265 |
| | | $\overline{LB}$ | 0.2101 | 0.1066 | 0.1082 | 0.2272 | 0.2522 | 0.1761 |
| | | $\overline{UB}$ | 1.6754 | 1.8484 | 1.8189 | 1.7449 | 1.6998 | 1.7122 |
| | 100 | $\bar{\hat{\beta}}$ | 0.9396 | 0.8564 | 0.7950 | 0.9175 | 0.9669 | 0.9664 |
| | | $MSE$ | 12.168 | 14.440 | 13.274 | 6.2455 | 2.1611 | 2.1652 |
| | | $\overline{SE}$ | 0.0466 | 0.0833 | 0.0546 | 0.0299 | 0.0066 | 0.0133 |
| | | $\overline{LB}$ | 0.2426 | 0.1391 | 0.1407 | 0.2597 | 0.2847 | 0.2086 |
| | | $\overline{UB}$ | 1.6439 | 1.8169 | 1.7874 | 1.7134 | 1.6683 | 1.6807 |
| | 300 | $\bar{\hat{\beta}}$ | 1.0211 | 0.9379 | 0.8765 | 0.9990 | 1.0484 | 1.0479 |
| | | $MSE$ | 10.306 | 12.578 | 11.412 | 4.3835 | 0.2991 | 0.3032 |
| | | $\overline{SE}$ | 0.0334 | 0.0701 | 0.0414 | 0.0167 | 0.0066 | 0.0001 |
| | | $\overline{LB}$ | 0.2751 | 0.1716 | 0.1732 | 0.2922 | 0.3172 | 0.2411 |

| | | | $\overline{UB}$ | 1.6124 | 1.7854 | 1.7559 | 1.6819 | 1.6368 | 1.6492 |

**Table 14.** Parameter estimation results for $t_{10}$ (Case 3 AR order=1)

| | $N$ | | OLS | ridge | LASSO | SCAD | bridge | elastic-net |
|---|---|---|---|---|---|---|---|---|
| $t_{10}$ | 50 | $\hat{\beta}_1$ | 5.6484 | 6.2749 | 6.2135 | 6.1360 | 6.0852 | 6.1749 |
| | | $MSE$ | 11.5626 | 11.7627 | 12.0461 | 11.7400 | 11.5755 | 11.5927 |
| | | $\overline{SE}$ | 2.9285 | 3.3648 | 3.9465 | 3.5818 | 2.8284 | 3.0652 |
| | | $\overline{LB}$ | 0.4101 | 0.1366 | -0.1918 | -0.0728 | 0.4521 | 0.3368 |
| | | $\overline{UB}$ | 6.0754 | 6.4514 | 6.4189 | 5.7449 | 5.6753 | 5.6484 |
| | | $\hat{\beta}_2$ | 4.2447 | 4.0017 | 4.0045 | 4.1201 | 4.1225 | 4.1640 |
| | | $MSE$ | 10.3697 | 10.4281 | 10.4270 | 10.4175 | 10.3623 | 10.3731 |
| | | $\overline{SE}$ | 2.2806 | 2.3684 | 2.3922 | 2.3617 | 2.2704 | 2.3728 |
| | | $\overline{LB}$ | -0.1096 | -0.2487 | -0.2955 | -0.0323 | -0.1156 | -0.1487 |
| | | $\overline{UB}$ | 3.2108 | 3.3417 | 3.3846 | 3.2522 | 3.2012 | 3.2418 |
| | | $\hat{\beta}_5$ | 5.0589 | 5.3039 | 5.4100 | 4.9008 | 4.8239 | 4.8477 |
| | | $MSE$ | 11.0493 | 11.1862 | 11.3226 | 11.1435 | 11.0786 | 11.0858 |
| | | $\overline{SE}$ | 3.1500 | 3.4283 | 3.3869 | 3.3617 | 3.2249 | 3.1918 |
| | | $\overline{LB}$ | -0.4734 | -0.6036 | -0.6426 | -0.5630 | -0.4826 | -0.4988 |
| | | $\overline{UB}$ | 4.6687 | 5.1499 | 5.2025 | 5.1645 | 4.7182 | 4.7497 |
| | 100 | $\hat{\beta}_1$ | 5.6664 | 5.5273 | 6.0484 | 6.0611 | 5.9031 | 6.0270 |
| | | $MSE$ | 9.8040 | 10.0135 | 10.3024 | 9.9943 | 9.8170 | 9.8147 |
| | | $\overline{SE}$ | 2.7948 | 3.2101 | 3.7513 | 3.3163 | 2.6864 | 2.8706 |
| | | $\overline{LB}$ | 0.5136 | 0.2208 | 0.0833 | 0.0007 | 0.5012 | 0.3951 |
| | | $\overline{UB}$ | 5.5931 | 6.2338 | 6.2536 | 5.7214 | 5.5130 | 5.8380 |
| | | $\hat{\beta}_2$ | 4.2906 | 4.0626 | 4.0983 | 4.1413 | 4.1776 | 4.1826 |
| | | $MSE$ | 9.6583 | 9.7179 | 9.7150 | 9.6994 | 9.6617 | 9.6679 |
| | | $\overline{SE}$ | 2.2090 | 2.2084 | 2.2914 | 2.2722 | 2.2670 | 2.2884 |
| | | $\overline{LB}$ | -0.0356 | -0.1789 | -0.1787 | -0.0973 | -0.0356 | -0.1529 |
| | | $\overline{UB}$ | 3.1196 | 3.2364 | 3.2778 | 3.2152 | 3.1796 | 3.1364 |
| | | $\hat{\beta}_5$ | 4.7341 | 5.2923 | 5.3911 | 4.8949 | 4.7711 | 4.8123 |
| | | $MSE$ | 10.1674 | 10.3430 | 10.5273 | 10.3513 | 10.3274 | 10.3646 |
| | | $\overline{SE}$ | 2.9928 | 3.0650 | 3.0635 | 3.2068 | 2.9828 | 3.0650 |
| | | $\overline{LB}$ | -0.1106 | -0.2510 | -0.2226 | -0.1394 | -0.0109 | -0.1507 |
| | | $\overline{UB}$ | 4.3300 | 4.7264 | 4.7048 | 4.6692 | 4.4300 | 4.4964 |
| | 300 | $\hat{\beta}_1$ | 5.9664 | 5.9273 | 5.9484 | 5.8611 | 5.8112 | 5.8273 |
| | | $MSE$ | 8.7040 | 8.7135 | 8.7324 | 8.7143 | 8.7120 | 8.7135 |
| | | $\overline{SE}$ | 2.0948 | 2.1801 | 2.1947 | 2.1163 | 2.1448 | 2.2101 |
| | | $\overline{LB}$ | 0.8136 | 0.3208 | 0.4833 | 0.3007 | 0.7131 | 0.5208 |
| | | $\overline{UB}$ | 5.3924 | 5.8338 | 5.9136 | 5.5214 | 5.4931 | 5.5128 |
| | | $\hat{\beta}_2$ | 4.3474 | 4.1194 | 4.1551 | 4.1981 | 4.2344 | 4.2394 |
| | | $MSE$ | 8.6751 | 8.7011 | 8.6982 | 8.6826 | 8.6691 | 8.6629 |
| | | $\overline{SE}$ | 2.1922 | 2.1916 | 2.2746 | 2.2554 | 2.2502 | 2.2716 |
| | | $\overline{LB}$ | 0.0212 | -0.1221 | -0.1219 | -0.0405 | 0.0212 | -0.0961 |
| | | $\overline{UB}$ | 3.0628 | 3.1796 | 3.2210 | 3.1584 | 3.1228 | 3.0796 |

|  |  |  | OLS | ridge | LASSO | SCAD | bridge | elastic-net |
|---|---|---|---|---|---|---|---|---|
|  |  | $\hat{\beta}_5$ | 4.7909 | 5.2355 | 5.3343 | 4.8381 | 4.7143 | 4.7555 |
|  |  | $MSE$ | 9.1506 | 9.3262 | 9.5105 | 9.3345 | 9.3106 | 9.3478 |
|  |  | $\overline{SE}$ | 2.9760 | 3.0482 | 3.0467 | 3.1900 | 2.9660 | 3.0482 |
|  |  | $\overline{LB}$ | -0.0538 | -0.1942 | -0.1658 | -0.0826 | 0.0459 | -0.0939 |
|  |  | $\overline{UB}$ | 4.2732 | 4.6696 | 4.6480 | 4.6124 | 4.3732 | 4.4396 |

**Table 15**. Parameter estimation results for $t_3$ (Case 3 AR order=1)

|  | $N$ |  | OLS | ridge | LASSO | SCAD | bridge | elastic-net |
|---|---|---|---|---|---|---|---|---|
| $t_3$ | 50 | $\hat{\beta}_1$ | 4.5669 | 4.3564 | 4.2950 | 4.2175 | 4.1667 | 4.2564 |
|  |  | $MSE$ | 4.8871 | 4.7872 | 5.0706 | 4.7645 | 4.6000 | 4.6172 |
|  |  | $\overline{SE}$ | 0.9530 | 1.3893 | 1.9710 | 1.6063 | 0.8529 | 1.0897 |
|  |  | $\overline{LB}$ | 0.9776 | 0.7041 | 0.3757 | 0.4947 | 1.0196 | 0.9043 |
|  |  | $\overline{UB}$ | 5.4069 | 5.7829 | 5.7504 | 5.0764 | 5.0068 | 4.9799 |
|  |  | $\hat{\beta}_2$ | 2.1632 | 2.0832 | 2.0860 | 2.2016 | 2.2040 | 2.2455 |
|  |  | $MSE$ | 2.3942 | 2.4526 | 2.4515 | 2.4420 | 2.3868 | 2.3976 |
|  |  | $\overline{SE}$ | 0.3051 | 0.3929 | 0.4167 | 0.3862 | 0.2949 | 0.3973 |
|  |  | $\overline{LB}$ | 0.4579 | 0.3188 | 0.2720 | 0.5352 | 0.4519 | 0.4188 |
|  |  | $\overline{UB}$ | 2.5423 | 2.6732 | 2.7161 | 2.5837 | 2.5327 | 2.5733 |
|  |  | $\hat{\beta}_5$ | 2.9774 | 3.3854 | 3.4915 | 2.9823 | 2.9054 | 2.9292 |
|  |  | $MSE$ | 3.0738 | 3.2107 | 3.3471 | 3.1680 | 3.1031 | 3.1103 |
|  |  | $\overline{SE}$ | 1.1745 | 1.4528 | 1.4114 | 1.3862 | 1.2494 | 1.2163 |
|  |  | $\overline{LB}$ | 0.0941 | -0.0361 | -0.0751 | 0.0045 | 0.0849 | 0.0687 |
|  |  | $\overline{UB}$ | 4.0002 | 4.4814 | 4.5340 | 4.4960 | 4.0497 | 4.0812 |
|  | 100 | $\hat{\beta}_1$ | 3.5849 | 3.6088 | 4.1299 | 4.1426 | 3.9846 | 4.1085 |
|  |  | $MSE$ | 2.9285 | 2.7380 | 3.0269 | 2.7188 | 2.5415 | 2.5392 |
|  |  | $\overline{SE}$ | 0.8193 | 1.2346 | 1.7758 | 1.3408 | 0.7109 | 0.8951 |
|  |  | $\overline{LB}$ | 1.0811 | 0.7883 | 0.6508 | 0.5682 | 1.0687 | 0.9626 |
|  |  | $\overline{UB}$ | 4.9246 | 5.5653 | 5.5851 | 5.0529 | 4.8445 | 5.1695 |
|  |  | $\hat{\beta}_2$ | 2.2091 | 2.1441 | 2.1798 | 2.2228 | 2.2591 | 2.2641 |
|  |  | $MSE$ | 2.3802 | 2.4424 | 2.4395 | 2.4239 | 2.3862 | 2.3924 |
|  |  | $\overline{SE}$ | 0.2335 | 0.2329 | 0.3159 | 0.2967 | 0.2915 | 0.3129 |
|  |  | $\overline{LB}$ | 0.5319 | 0.3886 | 0.3888 | 0.4702 | 0.5319 | 0.4146 |
|  |  | $\overline{UB}$ | 2.4511 | 2.5679 | 2.6093 | 2.5467 | 2.5111 | 2.4679 |
|  |  | $\hat{\beta}_5$ | 2.6526 | 3.3738 | 3.4726 | 2.9764 | 2.8526 | 2.8938 |
|  |  | $MSE$ | 2.8919 | 3.0675 | 3.2518 | 3.0758 | 3.0519 | 3.0891 |
|  |  | $\overline{SE}$ | 1.0173 | 1.0895 | 1.0880 | 1.2313 | 1.0073 | 1.0895 |
|  |  | $\overline{LB}$ | 0.4569 | 0.3165 | 0.3449 | 0.4281 | 0.5566 | 0.4168 |
|  |  | $\overline{UB}$ | 3.6615 | 4.0579 | 4.0363 | 4.0007 | 3.7615 | 3.8279 |
|  | 300 | $\hat{\beta}_1$ | 3.8849 | 4.0088 | 4.0299 | 3.9426 | 3.8927 | 3.9088 |
|  |  | $MSE$ | 2.4285 | 2.4380 | 2.4569 | 2.4388 | 2.4365 | 2.4380 |
|  |  | $\overline{SE}$ | 0.1193 | 0.2046 | 0.2192 | 0.1408 | 0.1693 | 0.2346 |
|  |  | $\overline{LB}$ | 1.3811 | 0.8883 | 1.0508 | 0.8682 | 1.2806 | 1.0883 |

| | | | | | | |
|---|---|---|---|---|---|---|
| $\overline{UB}$ | 4.7239 | 5.1653 | 5.2451 | 4.8529 | 4.8246 | 4.8443 |
| $\hat{\beta}_2$ | 2.2659 | 2.2009 | 2.2366 | 2.2796 | 2.3159 | 2.3209 |
| $MSE$ | 2.3634 | 2.4256 | 2.4227 | 2.4071 | 2.3694 | 2.3756 |
| $\overline{SE}$ | 0.2167 | 0.2161 | 0.2991 | 0.2799 | 0.2747 | 0.2961 |
| $\overline{LB}$ | 0.5887 | 0.4454 | 0.4456 | 0.5270 | 0.5887 | 0.4714 |
| $\overline{UB}$ | 2.3943 | 2.5111 | 2.5525 | 2.4899 | 2.4543 | 2.4111 |
| $\hat{\beta}_5$ | 2.7094 | 3.3170 | 3.4158 | 2.9196 | 2.7958 | 2.8370 |
| $MSE$ | 2.8751 | 3.0507 | 3.2350 | 3.0590 | 3.0351 | 3.0723 |
| $\overline{SE}$ | 1.0005 | 1.0727 | 1.0712 | 1.2145 | 0.9905 | 1.0727 |
| $\overline{LB}$ | 0.5137 | 0.3733 | 0.4017 | 0.4849 | 0.6134 | 0.4736 |
| $\overline{UB}$ | 3.6047 | 4.0011 | 3.9795 | 3.9439 | 3.7047 | 3.7711 |

## 5. Real data example

In this section, the data set which is used by Ramanathan (1998) is discussed to compare the proposed methods under $t$ distribution assumption with normal distribution assumption. We discussed the data set in terms of parameter estimation and variable selection cases. The data set considers the electricity consumption provided to the households by San Diago Electric and Gas Company. Data consisted of 87 quarters between 1972 and 1993. The response variable is the logarithmic value of the electricity consumed by the houses in kilowatts (LKWH). The explanatory variables are incomes of householders (LY), electricity kilowatt price (LPRICE), the energy required to cool houses (CDD) and the energy required to heat houses (HDD). The regression model obtained by Ramanathan (1998) and the expected parameter signs are as follows.

$$LKWH = \beta_0 + \beta_1 LY + \beta_2 LPRICE + \beta_3 CDD + \beta_4 HDD + \varepsilon_t$$
$$\beta_1 > 0, \ \beta_2 < 0, \beta_3 > 0, \beta_4 > 0.$$

In case of omission of autoregressive error assumption, the sign of the parameter of housing income (LY) variable was obtained as negative while the positive result was expected. However, when the error terms were modeled with the autoregressive model, a 4th degree autoregressive model was obtained according to the ACF (autocorrelation function) and PACF (partial autocorrelation function) graphs, and under this assumption, as a result of the re-parameter estimations, the signs for all parameter estimates were obtained as expected. In this study, parameter estimations were re-made by using the proposed variable selection methods. The results of these estimates are given in Table 16. From this table we can see that if the autoregressive error term assumption is not considered although it exists, the expected sign of the LY variable is appeared in opposite way. From this result we can understand that the using of autoregressive error term assumption is crucial for this type of data. By using AR(4) assumption and $t$ distributed error term the performance of the variable selection methods are given in Table 17. According to this table OLS and ridge with AR(4) assumption select all the variables significant for the model. On the other hand, LASSO, SCAD and the bridge methods are finding the LY insignificant. When we model the data by using $t$ distribution with 3 degrees of freedom and 4th degree of autoregression, $t$-ridge, $t$-LASSO and $t$-SCAD methods select LY and LPRICE variables. However, $t$-bridge and $t$-elastic-net methods finds only LPRICE variable significant. From these results we can say that only the electricity kilowatt price is affected the amount of electricity consumption in the houses.

**Table 16.** Parameter estimation results for Electric and Gas Company data.

|        |              | OLS      | OLS(AR)  |
|--------|--------------|----------|----------|
| LY     | $\hat{\beta}_1$ | **-0.00234** | 0.18625  |
| LPRICE | $\hat{\beta}_2$ | -0.01856 | -0.09354 |
| CDD    | $\hat{\beta}_3$ | 0.06365  | 0.00029  |
| HDD    | $\hat{\beta}_4$ | 0.08564  | 0.00022  |
| AR order |            | -        | 4        |

**Table 17.** Selected variables for Electric and Gas Company data.

| Method | Selected variables |
|--------|--------------------|
| OLS | (1,2,3,4) |
| ridge | (1,2,3,4) |
| LASSO | (2,3,4) |
| SCAD | (2,3,4) |
| bridge ($\gamma = 0.7$) | (2,3,4) |
| elastic-net | (2,3,4) |
| $t_3$-ridge | (2,3,4) |
| $t_3$-LASSO | (1,2) |
| $t_3$-SCAD | (1,2) |
| $t_3$-bridge ($\gamma = 0.7$) | (2) |
| $t_3$-elastic-net | (2) |

## 6. Conclusion

We have proposed the *t* distributed variable selection methods with autoregressive error term regression models to improve the resistance of the mentioned methods against the outliers. The variable selection methods; LASSO, SCAD, bridge and elastic-net based on *t* distributed are robust alternatives to the normal ones. We have provided an ECM algorithm to compute the proposed methods and performed a simulation study and a real data example to show the performance of the proposed methods. The simulation study shows that *t* distributed methods have superior performance under the existence of the outliers among the other methods in terms of both parameter estimation and variable selection. In terms of the determining the correct AR order for the error model, LASSO, SCAD, bridge and the elastic-net methods have nearly the same performance. On the other hand, if the degree of autoregressive order is getting higher, than the bridge and the elastic-net methods perform better than the others. Using *t* distribution as an error term assumption can be extended to the case of heavy tailed skew distributions. This problem will be a further study.

**Acknowledgments**

This study is a part of PhD dissertation in Ankara University, Graduate School of Natural and Applied Sciences.

# References

Alpuim, T. and El-Shaarawi, A., 2008. On the efficiency of regression analysis with AR(p) errors. Journal of Applied Statistics, 35:7, 717-737.

Arslan, O. 2016 Penalized MM regression estimation with Lγ penalty: a robust version of bridge regression, Statistics, 50:6.

Beach, C.M. and Mackinnon, J. G., 1978. A Maximum Likelihood Procedure for Regression with Autocorrelated Errors. Econometrica, vol. 46, no: 1, pp. 51- 58.

Box, E. P. G and Jenkins, M. G., 1976. Time Series Analysis: Forecasting and Control. Holden-Day: San Fransisco.

Dempster, A.P., Laird, N.M. and Rubin, D.B. 1977. Maximum likelihood from incomplete data via the EM algorithm. Journal of the Royal Statistical Society, Series B, 39,1-38.

Fun, J. and Li, R. 2001 Variable Selection via nonconcave penalized likelihood and its oracle properties. Journal of American Statistical Assoc. 96. 1348-1360.

Frank, I.E. and Friedman, J. (1993). A statistical view of some chemometrics regression tools. Thechnometrics, Vol. 35, pp. 109-148.

Hoerl, A.E., and Kennard, R.W., 1970 ridge regression: Biased estimation for nonorthagonal problems. Technometrics.

Lange, K.L., Little, R.J.A. and J.M.G. Taylor, 1989. *Robust statistical modeling using the t-distribution*, J. Am. Stat. Assoc. 84 pp. 881–896.

Liu, Y., Zhang, H.H., Park, C., and Ahn, J. 2007. Support vector machines with adaptive Lq penalty. Computational Statistics & Data Analysis, Vol. 51, pp.6380-6394.

Lucas, A. 1997. Robustness of the student *t-based* M-estimator. Communications in Statistics – Theory and Methods, 26(5), pp: 1165-1182.

McLachilan, G. J., Krishnan, T., (1997) The EM Algorithm and Extensions, Wiley series in probability and statistics, USA.

Meng, X. L. and Rubin, D. B. (1993). Maximum likelihood estimation via the ECM algorithm: A general framework. Biometrika 80, 267-278

Park, C., and Yoon, Y. J. 2011. bridge regression: Adaptivity and group selection.

Journal of Statistical Planning and Inference, Vol. 141, pp. 3506-3519.

Tibshiran, R. (1996). Regression shrinkage and selection via the LASSO. Journal of the Royal Statistical Society B, Vol. 58, pp. 267-288.

Tiku, M., Wong, W., & Bian, G., 2007, Estimating Parameters In Autoregressive Models In Non-Normal Situations: Symmetric Innovations . Communications in Statistics - Theory and Methods, 28 (2), 315-341.

Tuaç Y., Güney Y., Şenoğlu B. and Arslan O., 2018 Robust parameter estimation of regression model with AR(p) error terms, Communications in Statistics – Simulation and Computation, 47:8, 2343-2359.

Tuaç, Y., Güney, Y., & Arslan, O. (2020). Parameter estimation of regression model with AR (p) error terms based on skew distributions with EM algorithm. Soft Computing, 24(5), 3309-3330.

Tuaç, Y. 2020, Robust Parameter Estımatıon and Model Selectıon in Autoregressıve Error Term Regressıon Modelsi, Ph.D dissertion, Ankara University, Graduate School of Natural and Applied Sciences, Ankara.

Wang, H., Guodong L., and Chih-L,T., 2007. 'Regression Coefficient and Autoregressive Order Shrinkage and Selection via the Lasso'. *Journal of the Royal Statistical Society: Series B (Statistical Methodology)* 69, no. 1

Yoon, J.Y., Park, C. and Lee, T., 2012. Penalized Regression Models with Autoregressive Error Terms, Journal of Statistical Computation and Simulation, 83:9,1756-1772.

Zou, H. and Hastie, T., 2005 Regularization and variable selection via the elastic net. J.R. Statist. Soc. B. 67, part 2, pp. 301-320.